\documentclass[aps,amsmath,showpacs,amsfonts,10pt]{revtex4}
\usepackage{epsfig,graphicx}
\usepackage[english]{babel}
\usepackage{amsfonts}
\usepackage{amsmath}
\usepackage{latexsym}
\usepackage{graphics,bm}
\usepackage{natbib}
\usepackage{dcolumn}
\usepackage{bm}
\usepackage{rotating}

\begin{document}

\title{Optomechanical effect on the Dicke quantum phase transition and quasi-particle damping in a Bose-Einstein Condensate: A new tool to measure weak force}

\author{Neha Aggarwal$^{1,2}$ and Aranya B Bhattacherjee$^{2,3}$}

\address{$^{1}$Department of Physics and Astrophysics, University of Delhi, Delhi-110007, India} \address{$^{2}$Department of Physics, ARSD College, University of Delhi (South Campus), New Delhi-110021, India}\address{$^{3}$School of Physical Sciences, Jawaharlal Nehru University, New Delhi-110067, India}

\begin{abstract}
We make a semi-classical steady state analysis of the influence of mirror motion on the quantum phase transition for an optomechanical Dicke model in the thermodynamic limit. An additional external mechanical pump is shown to modify the critical value of atom-photon coupling needed to observe the quantum phase transition. We further show how to choose the mechanical pump frequency and cavity-laser detuning to produce extremely cold condensates. The present system can be used as a quantum device to measure weak forces.

\end{abstract}


\maketitle

\section{Introduction}

In recent years, the underlying mechanism of radiation-pressure forces has experimentally entangled the two distinct subjects of nanomechanical resonators and optical microcavities. There has been a great surge of interest in the field of cavity optomechanics with a wide variety of systems, such as gravitational wave detectors \citep{1,2}, nanomechanical cantilevers \citep{3,4,5,6,7,8}, membranes \citep{9}, vibrating microtoroids \citep{10,11}, atomic ensembles \citep{12,13,14,15,16} and Bose-Einstein Condensates (BECs) \citep{17,18,aran,aran1,sonam,sonam1}. The quantum optical properties of mirror interacting with the cavity field via radiation pressure exhibit interesting similarities to an intracavity Kerr-like interaction \citep{19,20}. The optomechanical quantum ground state cooling of nano- and micro-mechanical oscillators is used in a large variety of sensitive measurements like detection of weak forces \citep{brag,abra,bocko}, small masses \citep{jensen} and small displacements \citep{lalt}. The dynamical instability of a driven cavity involving a movable mirror has been studied recently in the context of classical investigations of non-linear regimes \citep{21}. The coupled dynamics of atoms trapped within the optical cavity with movable mirror has also been studied \citep{22}.

BEC is considered as an ideal coherent system to illustrate the many-body quantum physics in a highly controllable manner. The breakthrough success of producing BEC atoms \citep{23,24,25,26} has created a new interesting era of investigative study \citep{27,28,29}. It is made possible due to the largely improved cooling techniques \citep{30}. The coupling of an atom with the electromagnetic field can be controlled precisely in the field of quantum electrodynamics \citep{31,32}. Thus, one can trap and cool the BEC atoms experimentally within an optical cavity \citep{33,34}. Moreover, the methods to observe the properties of ultracold atoms in the field of quantum degenerate gases are based on the matter-wave destructive interference between atoms released from traps \citep{35}. The superfluid-mott insulator phase transition is governed by short-range interactions \citep{35}. However, the phase transition induced by long-range interaction is the creation of a self-organized phase from a BEC in a high finesse optical cavity above a certain critical transverse optical pump intensity \citep{36}. The self-organization phenomenon has been investigated experimentally using a BEC in an optical cavity in recent years  \citep{37,38}. The spatial symmetry of the cavity optical lattice is spontaneously broken at the phase transition. In addition it was pointed out that this self organization transition is equivalent to the Dicke quantum phase transition \citep{37,40}.

The Dicke model \citep{40,41,42,43,44} describes the uniform interaction between the two-level atoms or spins with the light. The Dicke model exhibits a continuous phase transition to a state with a non-vanishing photon population when the atom-light coupling exceeds a certain critical value. In order to review the Dicke model and to study its applications in quantum optics, see Ref. \citep{45}.

Motivated by these interesting developments in the field of cavity optomechanics and ultracold gases, we propose an optomechanical system consisting of an elongated cigar-shaped two-level BEC interacting with a single mode of a high finesse optical cavity with one movable mirror. This optomechanical Dicke model is used to study the semi-classical steady states to make an analysis of quantum phase transition (self-organization process) in the thermodynamic limit. For the generalized Dicke model, a continuous phase transition is obtained on the basis of its semi-classical analysis, presented in Refs. \citep{37,38}. We also investigate the influence of cavity-mirror coupling on the self-organization of BEC in the optomechanical cavity.  We then further make a semi-classical steady state analysis of the influence of an external mechanical pump (external foce on the vibrating mirror) on the quantum phase transition. Changing the mechanical pump frequency exhibits a shift in the critical atom-photon coupling strength which is needed to observe the Dicke phase transition. We also show that this system can serve as a new quantum device to measure weak forces. Finally, we discuss how the presence of mechanical pump alters the damping of BEC confined within the optomechanical cavtiy. This is done by demonstrating the variation in condensate energy on changing the mechanical pump frequency for both the negative and positive cavity-laser detunings. Here, we combine the use of an optical cavity with the movable mirror excited by an external source i.e., a mechanical pump with the goal of producing colder condensates.

\section{Steady State Analysis for the Optomechanical Dicke Model}

In this section, we introduce and study the basic model for our system as shown in fig.(1). The optomechanical system considered here essentially involves a Fabry-Perot optical cavity with one fixed mirror and another movable mirror of mass $M$ oscillating freely at mechanical frequency $\omega_{m}$. We have in addition an elongated cigar shaped N two-level BEC atoms, with mass $m$ and transition frequency $\omega_{0}$. This atomic condensate is coupled to a single standing wave cavity mode of frequency $\omega_{c}$ and decay rate $\kappa$ of a high-finesse optical cavity of length $L$, and is driven by an external laser of frequency $\omega_{l}$ perpendicular to the cavity axis \citep{37}. We will consider the system dynamics in one dimension only i.e., along the axis of cavity for simplicity. A tight harmonic potential of frequency $\omega_{r}$ freezes out the radial motion of the BEC such that its spatial dimension along the cavity axis is taken into consideration only. We are assuming the atom-laser detuning $\Delta_{0} (=\omega_{0}-\omega_{l})$ to be very large such that the atomic spontaneous emission rate is negligible and the cavity photon loss will be the dominant dissipative process. The electronically excited atomic state can be adiabatically eliminated as it is justified for large detuning $\Delta_{0}$. As a result, an effective two-level atomic system with zero momentum state $|p>=|0>$ and excited momentum state $|p>=|\pm\hbar k>$ is formed, where p being the momenta along the cavity axis and $k$ denotes the wave vector of the pump laser field \citep{37,dimer}. These states are coupled through a pair of distinct Raman channels such that the effective transition frequency $\omega_{a}$ is twice the atomic recoil frequency $\omega_{R}=\hbar k^{2}/2m$. When all BEC atoms with these different momentum states are coupled identically with the single-mode cavity field, the simplest model of such system is provided by the optomechanical Dicke Hamiltonian given as \citep{aran,37,dimer}:

\begin{equation}\label{ham}
\hat{H}_{om}=\hbar\omega_{a}\hat{J}_{z}+\hbar\omega_{c}\hat{a}^{\dagger}\hat{a}+
\hbar\omega_{m}\hat{b}^{\dagger}\hat{b}+\hbar\omega_{c}\delta_{0}\hat{a}^{\dagger}\hat{a}(\hat{b}+\hat{b}^{\dagger})+\hbar\frac{\lambda}{\sqrt{N}}(\hat{a}+\hat{a}^{\dagger})(\hat{J}_{+}+\hat{J}_{-}),
\end{equation}

where $\hat{J}_{z}$, $\hat{J}_{+}$ and $\hat{J}_{-}$ are the collective atomic operators satisfying angular momentum commutation relations $[\hat{J}_{+},\hat{J}_{-}]=2\hat{J}_{z}$ and $[\hat{J}_{\pm},\hat{J}_{z}]=\mp \hat{J}_{\pm}$. They are expressed as $\hat{J_{+}}=\hat{J_{-}}^{\dagger}=\sum_{n}|\pm \hbar k>_{n}$ $_{n}<0|$ and $J_{z}=\sum_{n}(|\pm \hbar k>_{n}$ $_{n}<\pm \hbar k|$ $-|0>_{n}$ $_{n}<0|)$, where the index n labels the condensate atom. The collective atom-photon coupling strength is denoted by $\lambda$ which can be experimentally tuned by varying the pump laser power \citep{37}. The cavity field operators  ${\hat{a},\hat{a}^{\dagger}}$ represent the respective annihilation and creation operators following the commutation relation $[\hat{a},\hat{a}^{\dagger}]=1$. The annihilation (creation) operator of the mechanical oscillator is denoted by $\hat{b}$ $(\hat{b}^{\dagger})$ $([\hat{b},\hat{b}^{\dagger}]=1)$. The contact interactions between the atoms of the condensate is neglected here. The several mechanical degrees of freedom arising from the radiation pressure are neglected by using a band-pass filter in the detection scheme such that only a single vibrational mode is considered \citep{46}. The optomechanical interaction between the cavity field and the oscillating mirror is due to the pressure exerted by the intra-cavity photons on the mirror. Depending upon the number of photons in the cavity, the light field exerts a force on the oscillating mirror which shifts the phase of the field by $2k_{1}l_{m}$. Here, $l_{m}$ represents the mirror displacement from its equilibrium position and $k_{1}$ denotes the propagation wave vector of the cavity field. The nonlinear dispersive coupling between the intensity of the cavity field and the position quadrature of the movable mirror is represented by $\delta_{0}$ where $\delta_{0}<<1$.

\begin{figure}[h]
\hspace{-0.0cm}
\includegraphics [scale=0.8]{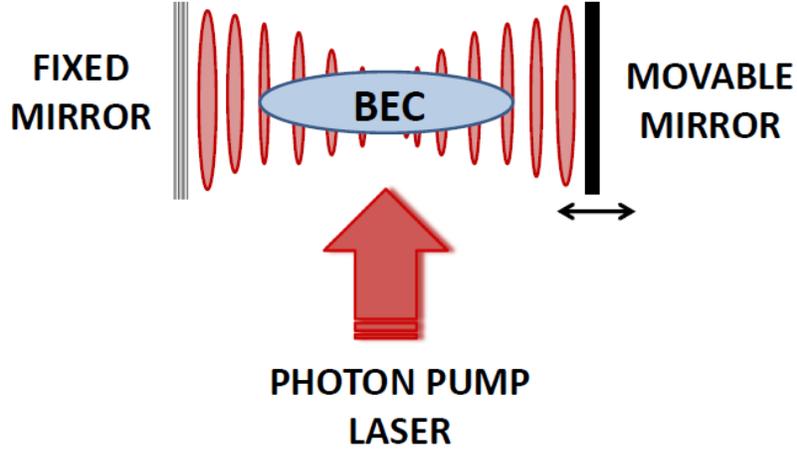}
\caption{(color online) Schematic representation of the optomechanical system we are investigating here. It involves the Bose-Einstein Condensate confined within a high-finesse optical cavity driven by a transverse pump laser. One of the cavity mirror is movable.}
\end{figure}\label{fig1}

\begin{figure}[h]
\hspace{-0.0cm}
\begin{tabular}{cc}
\includegraphics [scale=0.80]{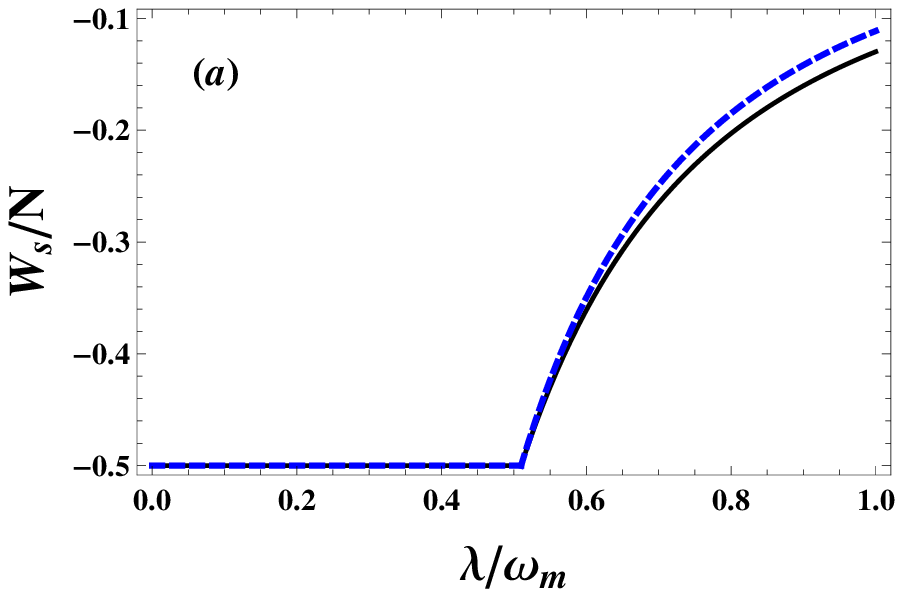}& \includegraphics [scale=0.80] {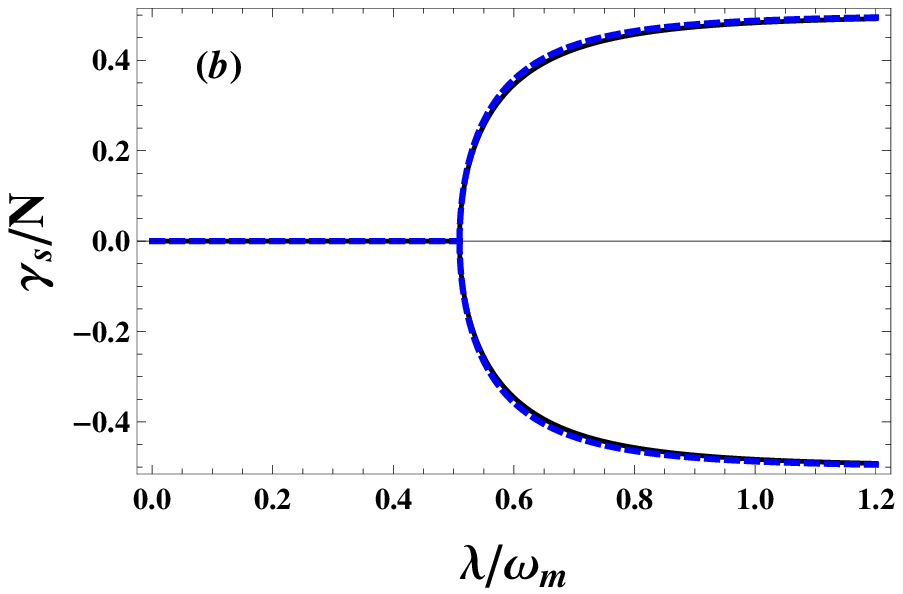}\\
\includegraphics [scale=0.80]{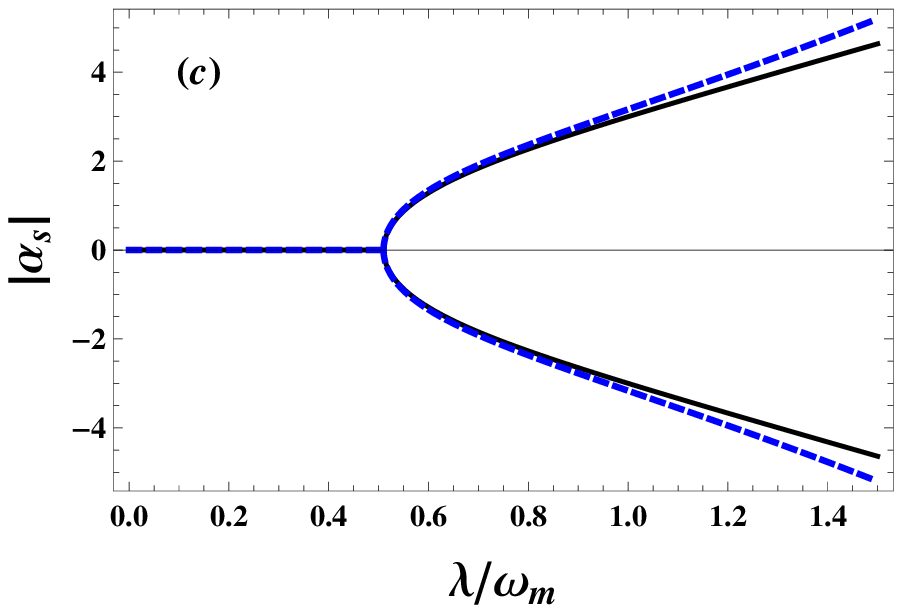}& \includegraphics [scale=0.80] {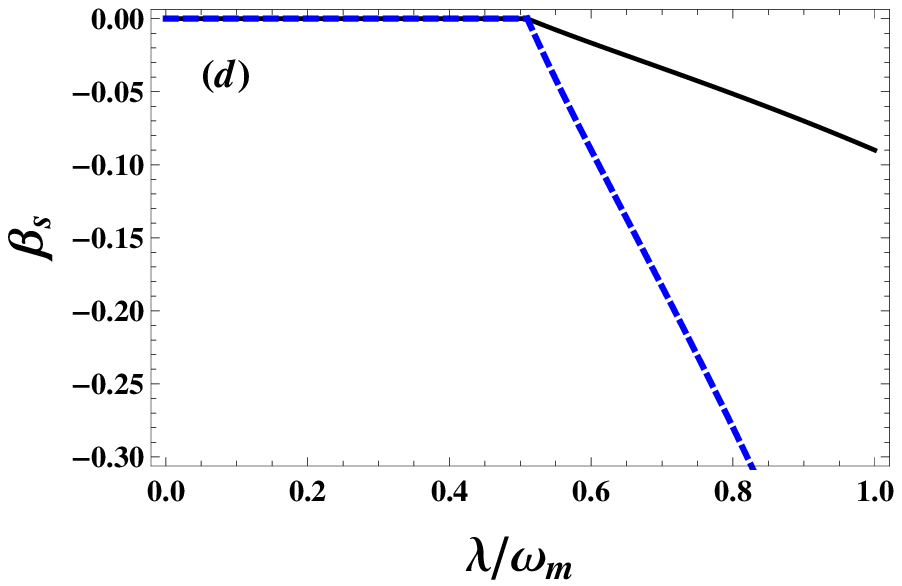}\\
\end{tabular}
\caption{(color online) Plot of steady state atomic population inversion $w_{s}$, polarization amplitude $\gamma_{s}$, absolute value of cavity field amplitude $\mid\alpha_{s}\mid$ and mirror mode amplitude $\beta_{s}$ as a function of dimensionless atom-photon coupling strength $\lambda/\omega_{m}$ for two different values of mirror-photon coupling $\delta_{0}=0.01$ (solid line) and $\delta_{0}=0.05$ (dashed line) with $\omega_{a}=\omega_{m}$ and $\omega_{c}=\omega_{m}$. Other parameters used are: $\Gamma=10^{-5}\omega_{m}$, $\kappa=0.2\omega_{m}$ and $N=10$. Only stable steady states are shown.}
\end{figure}\label{fig2}

We aim to make a semiclassical analysis of the above mentioned optomechanical Dicke model given by Hamiltonian (\ref{ham}) in the thermodynamic limit of $N>>1$. The mean-field analysis of the system  can be demonstrated by introducing the c-number variables $\alpha\equiv<\hat{a}>$, $\beta\equiv<\hat{b}>$, $w\equiv<\hat{J}_{z}>$ and $\gamma\equiv<\hat{J}_{-}>$ where $\alpha$, $\beta$ and $\gamma$ are the complex cavity field, mirror mode and the atomic polarization amplitudes respectively. Here $w$ represents the population inversion and is real. The entanglement between the atomic and photonic subsystems is neglected by the mean-field approximation which can affect the transient dynamics towards the steady state significantly \citep{47}. The semiclassical equations of motion for the system Hamiltonian (\ref{ham}) can be written as:

\begin{equation}\label{eq1}
\dot{\alpha}=-(\kappa+i\omega_{c})\alpha-i\omega_{c}\delta_{0}\alpha(\beta+\beta^{\ast})-i\frac{\lambda}{\sqrt{N}}(\gamma+\gamma^{\ast}),
\end{equation}

\begin{equation}\label{eq2}
\dot{\beta}=-(\Gamma+i\omega_{m})\beta-i\omega_{c}\delta_{0}\mid\alpha\mid^{2},
\end{equation}

\begin{equation}\label{eq3}
\dot{\gamma}=-i\omega_{a}\gamma+2i\frac{\lambda}{\sqrt{N}}(\alpha+\alpha^{\ast})w,
\end{equation}

\begin{equation}\label{eq4}
\dot{w}=i\frac{\lambda}{\sqrt{N}}(\alpha+\alpha^{\ast})(\gamma-\gamma^{\ast}),
\end{equation}

where $\Gamma$ denotes the damping rate of the mechanical mode which arises due to the interaction of the vibrating mirror with environment. The equations (\ref{eq1}-\ref{eq4})are obtained by neglecting the quantum fluctuations and by imposing the factorization
$<a(b+b^{\dagger})> \rightarrow <a><(b+b^{\dagger})>$,
$<(a+a^{\dagger})J_{z}> \rightarrow <(a+a^{\dagger})><J_{z}>$,
$<(a+a^{\dagger})(J_{-}-J_{+})> \rightarrow <(a+a^{\dagger})><(J_{-}-J_{+})>$.
The semiclassical equations have to follow the constraint that the magnitude of psuedo-angular momentum $w^{2}+\mid\gamma\mid^{2}=\frac{N^{2}}{4}$ is conserved. Using this conservation law, we find the steady state values for the different c-number variables by factorizing the non-linear algebraic equations and setting their time-derivatives to zero. The steady state solution displays a bifurcation point at $\lambda=\lambda_{c}$, given as

\begin{equation}
\lambda_{c}=\frac{1}{2}\sqrt{\frac{\omega_{a}}{\omega_{c}}(\kappa^{2}+\omega_{c}^{2})},
\end{equation}

where $\lambda_{c}$ represents the critical value of atom-cavity coupling strength. The steady state solutions for $\lambda<\lambda_{c}$ are  given as:

\begin{equation}
\alpha_{s}=\beta_{s}=\gamma_{s}=0, w_{s}=\pm\frac{N}{2}.
\end{equation}

The states with positive and negative population inversion are dynamically unstable and stable respectively. For $\lambda>\lambda_{c}$, these solutions become unstable and new sets of stable solutions appear. The stable solution for the steady state population inversion above the critical value is obtained by solving the following cubic equation:

\begin{eqnarray}
w_{s}^{3}\left[\frac{\lambda^{2}\delta_{0}^{2}\sigma(1-2\bar{\epsilon})}{N\lambda_{c}^{2}}\right] +w_{s}\left[1-\frac{N\lambda^{2}\delta_{0}^{2}\sigma(1-2\bar{\epsilon})}{4\lambda_{c}^{2}} \right]+\frac{N\lambda_{c}^{2}}{2\lambda^{2}}=0,
\end{eqnarray}

where $\bar{\epsilon}=\frac{\omega_{c}^{2}}{\kappa^{2}+\omega_{c}^{2}}$ and $\sigma=\frac{2\omega_{m}\omega_{a}}{\Gamma^{2}+\omega_{m}^{2}}$. The above equation is solved numerically using Mathematica 9.0. The other set of stable steady states above the critical value are given as follows:

\begin{eqnarray}
\gamma_{s}=\pm\sqrt{\frac{N^{2}}{4}-w_{s}^{2}},
\end{eqnarray}

\begin{eqnarray}
\mid\alpha_{s}\mid=\pm\left[\frac{N(\kappa^{2}+\omega_{c}^{2})}{4\lambda^{2}\gamma_{s}^{2}}-\frac{4\delta_{0}^{2}\omega_{m}\omega_{c}
\bar{\epsilon}}{(\Gamma^{2}+\omega_{m}^{2})}\right]^{-1/2},
\end{eqnarray}

\begin{eqnarray}
\beta_{s}=\frac{-\omega_{c}\delta_{0}\mid\alpha_{s}\mid^{2}(\omega_{m}+i\Gamma)}{\Gamma^{2}+\omega_{m}^{2}}.
\end{eqnarray}

Fig.(2) depicts the plot of steady state atomic inversion $w_{s}$, polarization amplitude $\gamma_{s}$, absolute value of cavity field amplitude $\mid\alpha_{s}\mid$ and mirror mode amplitude $\beta_{s}$ as a function of dimensionless atom-cavity field coupling strength $\lambda/\omega_{m}$ for two different values of mirror-photon coupling with $\delta_{0}=0.01$ (solid line) and $\delta_{0}=0.05$ (dashed line). From this figure, note the bifurcation to states of finite amplitude and inversion as the coupling approaches the critical value $\lambda_{c}$. Fig.2(a) depicts an abrupt increase in the steady state atomic population inversion at the critical value of atom-photon coupling. Further note that $\gamma_{s}$ displays a similar kind of behaviour as $\mid\alpha_{s}\mid$ (see figs.2(b) and 2(c)). Fig.2(d) shows an abrupt increase in the steady state amplitude of the oscillating mirror at the critical atom-photon coupling value. The behaviour of steady-state semi-classical solutions above and below the bifurcation point demonstrates the existence of quantum phase transition (onset of self-organization of BEC in an optomechanical cavity). In the self organization process, there is an abrupt change in the steady state at a critical atom-photon coupling strength $\lambda_{c}$. In several earlier works, the Dicke Hamiltonian undergoing QPT has been studied \citep{40,42,43,44,48,49,50,51}. Thus the present result represents the Optomechanical Dicke model quantum phase transition in the absence of quantum fluctuations in the thermodynamic limit. Furthermore, the steady state values of the atomic population inversion, polarization amplitude, cavity photon amplitude and mirror mode amplitude increase with the increase in mirror-cavity field coupling above the critical point. An increase in steady state absolute value of cavity field amplitude due to the increase in mirror-photon coupling naturally leads to an increase in the radiation pressure. This in turn increases the steady state mechanical field amplitude  of the vibrating mirror. Moreover, the increase in $w_{s}$ with increase in $\delta_{0}$ illustrates the fact that the final phase of BEC confined in an optomechanical cavity is more organized for a higher mirror-photon coupling. Thus, the continuous monitoring of these steady state stable solutions could serve as an important tool to observe the Dicke phase transition.

In the next section, we make a theoretical analysis of the optomechanical Dicke model for the steady states in the presence of an external mechanical pump.

\section{Steady State Analysis for the Optomechanical Dicke Model in the presence of mechanical Pump}

In this section, we consider an optomechanical system in the presence of an external mechanical pump as shown in fig(3). This external source can be any mechanical object in physical contact with the movable mirror or an external laser that helps in oscillating the mirror via radiation pressure. We present here the steady state analysis of the semiclassical equations of motion for the optomechanical Dicke Hamiltonian involving an external mechanical pump. The pump excites the mirror by coupling with the amplitude quadrature of the mirror fluctuations. Thus, the Hamiltonian for the system considered here can be rewritten as \citep{aran,37,dimer,52}:

\begin{equation}\label{ham1}
\hat{H}_{omp}=\hbar\omega_{a}\hat{J}_{z}+\hbar\omega_{c}\hat{a}^{\dagger}\hat{a}+
\hbar\omega_{m}\hat{b}^{\dagger}\hat{b}+\hbar\omega_{c}\delta_{0}\hat{a}^{\dagger}\hat{a}(\hat{b}+\hat{b}^{\dagger})+\hbar\frac{\lambda}{\sqrt{N}}(\hat{a}+\hat{a}^{\dagger})(\hat{J}_{+}+\hat{J}_{-})+\hbar\eta_{p}(\hat{b}+\hat{b}^{\dagger}),
\end{equation}

\begin{figure}[h]
\hspace{-0.0cm}
\includegraphics [scale=0.8]{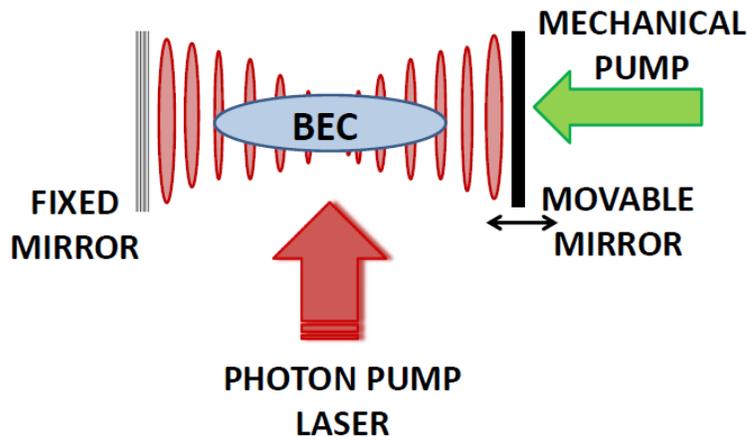}
\caption{(color online)setup of the model we are analysing here. The configuration is same as the fig(1) with an additional mechanical pump which further influences the mirror motion.}
\label{fig3}
\end{figure}

\begin{figure}[h]
\hspace{-0.0cm}
\begin{tabular}{cc}
\includegraphics [scale=0.80]{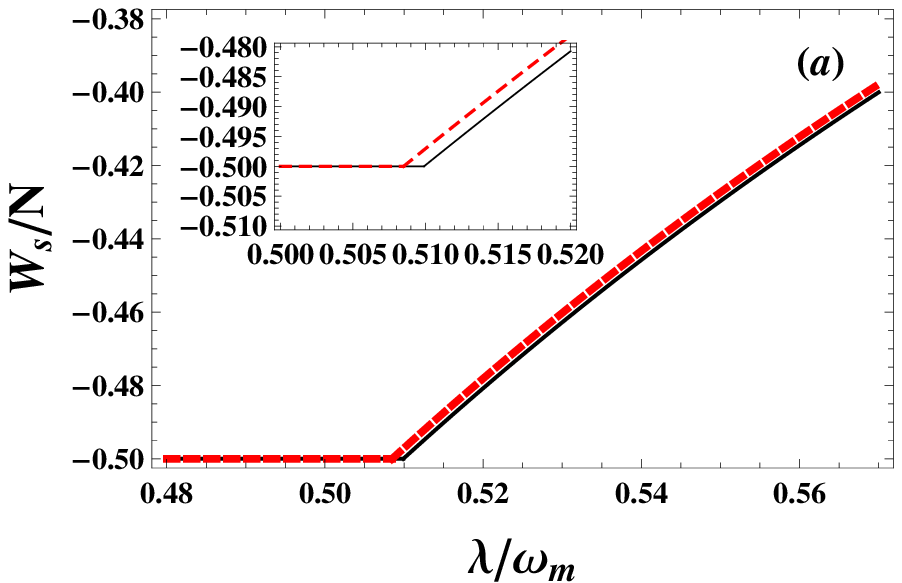}& \includegraphics [scale=0.80] {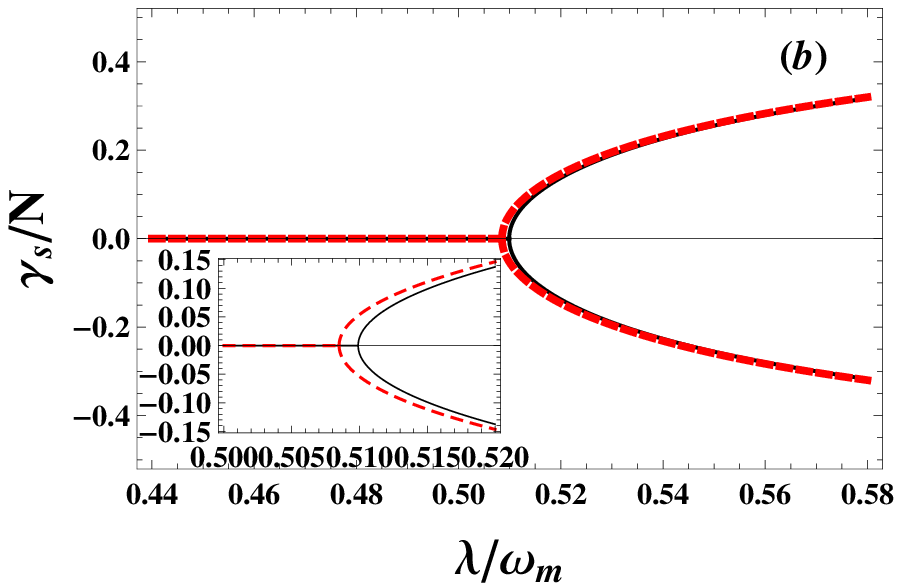}\\
\includegraphics [scale=0.80]{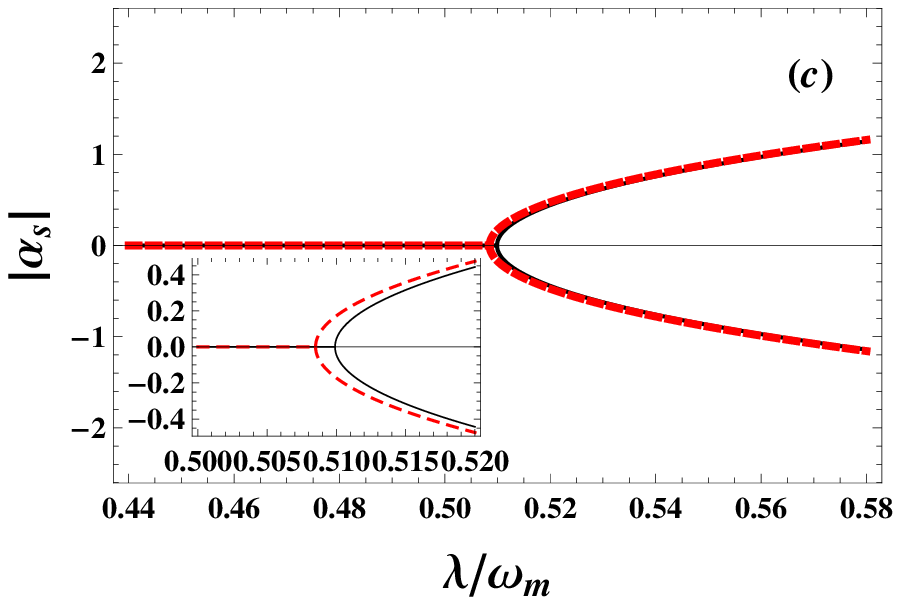}& \includegraphics [scale=0.80] {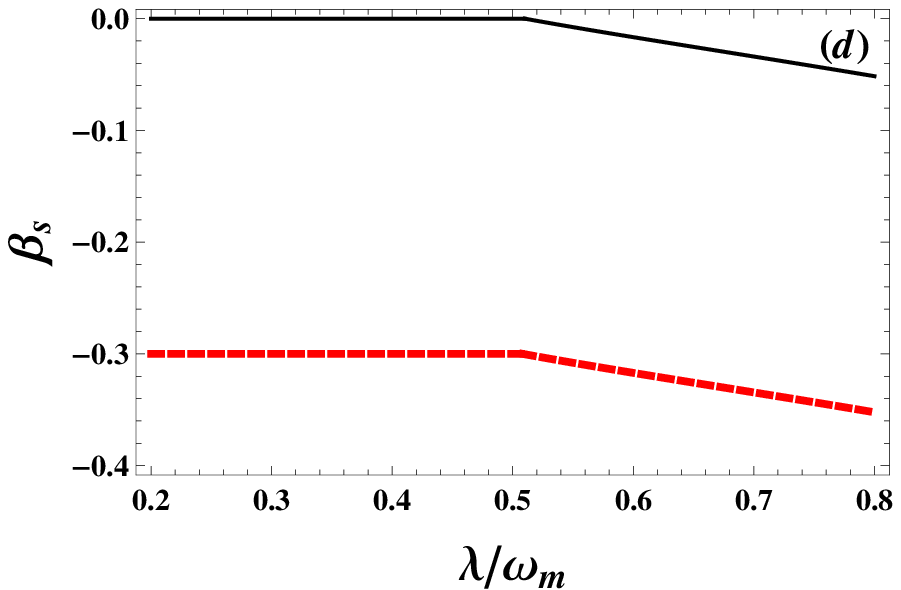}\\
\end{tabular}
\caption{(color online) Plot of steady state atomic population inversion $w_{s}$, polarization amplitude $\gamma_{s}$, absolute value of cavity field amplitude $\mid\alpha_{s}\mid$ and mirror mode amplitude $\beta_{s}$ as a function of dimensionless atom-cavity coupling strength $\lambda/\omega_{m}$ for $\delta_{0}=0.01$, $\omega_{a}=\omega_{m}$ and $\omega_{c}=\omega_{m}$ in the absence of mechanical pump (solid line) and in the presence of a mechanical pump having frequency $\eta_{p}=0.3\omega_{m}$ (dashed line). Other parameters used are the same as in fig(2). Only stable steady states are shown.}
\end{figure}\label{fig4}

Last term in the Hamiltonian represents the energy due to an external mechanical pump where $\eta_{p}$ represents the mechanical pump frequency which is considered to be small here i.e., $\eta_{p}<<1$. The other parameters of the system remain same. The semiclassical equation of motion for the mirror in the presence of mechanical pump is rewritten as:

\begin{equation}\label{eq5}
\dot{\beta}=-(\Gamma+i\omega_{m})\beta-i\omega_{c}\delta_{0}\mid\alpha\mid^{2}-i\eta_{p}.
\end{equation}

Under the constraint of conserved psuedo-angular momentum, we solve the equations (\ref{eq1}), (\ref{eq3}), (\ref{eq4}) and (\ref{eq5}) for the steady states, where the critical value of the atom-photon coupling strength is modified as:

\begin{eqnarray}
\lambda'_{c}=\frac{\lambda_{c}}{\sqrt{1-\frac{\sigma\eta_{p}\delta_{0}(1-2\bar{\epsilon})}{\omega_{a}}}}.
\end{eqnarray}

This clearly represents a shift in the bifurcation point from $\lambda_{c}$ to $\lambda'_{c}$ in the presence of mechanical pump. It, thereby, changes the expressions for steady states immediately. For $\lambda<\lambda'_{c}$, the steady states are given as:

\begin{equation}
\alpha_{s}=\gamma_{s}=0,\beta_{s}=\frac{-\eta_{p}(\omega_{m}+i\Gamma)}{\Gamma^{2}+\omega_{m}^{2}}, w_{s}=\pm\frac{N}{2},
\end{equation}

where the states having negative and positive population inversion are dynamically stable and unstable respectively. The states become unstable above the critical point. The stable steady state value of population inversion above the critical point is evaluated from the equation:

\begin{eqnarray}\label{Ws}
w_{s}^{3}\left[\frac{\delta_{0}^{2}\sigma\lambda^{2}(1-2\bar{\epsilon})}{N {\lambda'_{c}}^{2} X_{1}^{2}} \right]+w_{s}\left[1-\frac{\delta_{0}^{2}\sigma N
\lambda^{2}(1-2\bar{\epsilon})}{4{\lambda'_{c}}^{2}X_{1}^{2}} \right]+\frac{N{\lambda'_{c}}^{2}}{2\lambda^{2}}=0,
\end{eqnarray}

where,

\begin{equation}
X_{1}=1-\frac{\sigma\eta_{p}\delta_{0}(1-2\bar{\epsilon})}{\omega_{a}}.
\end{equation}

Eqn.(\ref{Ws}) has been solved numerically using Mathematica 9.0. The other modified stable steady state solutions above the bifurcation point are as follows:

\begin{eqnarray}
\gamma_{s}=\pm\sqrt{\frac{N^{2}}{4}-w_{s}^{2}},
\end{eqnarray}

\begin{eqnarray}
\mid\alpha_{s}\mid=\pm \frac{X_{2}}{X_{3}},
\end{eqnarray}

\begin{eqnarray}
\beta_{s}=\frac{-(\omega_{m}+i\Gamma)(\eta_{p}+\omega_{c}\delta_{0}\mid\alpha_{s}\mid^{2})}{\Gamma^{2}+\omega_{m}^{2}},
\end{eqnarray}

where,

\begin{equation}
X_{2}=\left[1+\frac{4\delta_{0}\omega_{m}\bar{\epsilon}\eta_{p}}{(\Gamma^{2}+\omega_{m}^{2})} \right]^{1/2},
\end{equation}

\begin{equation}
X_{3}=\left[\frac{N(\kappa^{2}+\omega_{c}^{2})}{4\lambda^{2}\gamma_{s}^{2}}-\frac{4\omega_{m}\omega_{c}\delta_{0}^{2}\bar{\epsilon}}{\Gamma^{2}+\omega_{m}^{2}} \right]^{1/2}.
\end{equation}

Fig.(4) represents the plot of steady state atomic inversion $w_{s}$, polarization amplitude $\gamma_{s}$, absolute value of cavity field amplitude $\mid\alpha_{s}\mid$ and amplitude of the vibrating mirror $\beta_{s}$ as a function of dimensionless atom-cavity field coupling strength $\lambda/\omega_{m}$ in the absence of mechanical pump (solid line) and in the presence of mechanical pump having frequency $\eta_{p}=0.3\omega_{m}$ (dashed line). It illustrates the influence of an external mechanical pump (external force on the oscillating mirror) on the quantum phase transition. The important point here is the shift in bifurcation point $\lambda_{c}$ in the presence of mechanical pump. We can infer from the figure that the additional mechanical pump shifts the phase transition at lesser value of atom-photon coupling. The external mechanical pump decreases the cavity length. Since the effective confinement region decreases, the effective optical potential increases. Hence a smaller value of atom-light coupling is required to observe the Dicke phase transiton. This implies that the phase transition point can be coherently controlled by an external mechanical pump. The oscillating mirror usually behaves as a ponder-motive detector in order to measure the weak forces acting on it \citep{vitali}. The mechanical pump considered here could be a weak force. Thus by appropriately calibrating the device, one can measure the  weak forces from the position of the critical value of atom-photon coupling.

The experimental observation of the optomechanical Dicke phase transition crucially depends on the loss of BEC atoms due to the mirror motion. Hence, in order to check the heating effect of BEC, we will study the influence of an external mechanical pump on the damping of condensate atoms held within the optomechanical cavity in the next section.

\section{Quasiparticle Damping in a Bose-Einstein Condensate}

In this section, we monitor the damping of Bose-Einstein Condensate, confined in a harmonic potential of frequency $\omega$ inside the optomechanical cavity involving an external mechanical pump, via energy of the condensate. The schematic representation of the system is depicted in fig(5). Some of the parameters in the optomechanical system considered here are different from the parameters defined previously. As shown in model figure, the system involves a BEC held with in an optomechanical cavity, driven by a separate laser $\Omega$ of frequency $\omega_{L}$ with wave vector $k_{L}$ and a pump laser in the presence of an external mechanical pump. The atom-laser detuning $\Delta_{a}=(\omega_{0}-\omega_{L})$ is kept large to suppress the spontaneous emission of photons by the condensate atoms as it can eventually destroy the condensate by producing heat. The other parameters used are same as defined in the previous section. The effective single-particle Hamiltonian in the dipole and rotating wave approximation for the system under consideration is reconstructed as \citep{52,53}:

\begin{equation}\label{ham2}
\hat{H}_{0}=\frac{\hat{p}^{2}}{2m}+\frac{m\omega^{2}\hat{x}^{2}}{2}+\hbar(\Delta-i\kappa)\hat{a}^{\dagger}\hat{a}+\hbar\omega_{m}\hat{b}^{\dagger}\hat{b}+\hbar[h(x)\hat{a}
+h^{*}(x)\hat{a}^{\dagger}]+\hbar \epsilon \omega_{m} \hat{a}^{\dagger}\hat{a}(\hat{b}+\hat{b}^{\dagger})+\hbar \eta_{p}(\hat{b}+\hat{b}^{\dagger}),
\end{equation}

where, the first term in the Hamiltonian represents the kinetic energy of the condensate atoms. Second term gives the potential energy of the atoms. Third term describes the energy of the cavity mode where $\hat{a}(\hat{a}^{\dagger})$ is the annihilation (creation) operator of the cavity mode such that $[\hat{a},\hat{a}^{\dagger}]=1$ and $\Delta(=\omega_{c}-\omega_{L})$ is the cavity-laser detuning. The fourth term represents the energy of the single mechanical mode of the mirror with $\hat{b}(\hat{b}^{\dagger})$ as the annihilation (creation) operator such that $[\hat{b},\hat{b}^{\dagger}]=1$. Fifth term illustrates the interaction between the condensate field and the light field such that $h(x)=-\Omega(x)g \cos(k'x)\Delta_{a}/(\Delta_{a}^{2}+\gamma^{2})$. Here, $k'$ is the wave vector of pump laser field. Also, $\gamma$ is the rate of atomic spontaneous emission and $\Omega(x)$ is assumed to be directly proportional to $e^{ik_{L}x}$. The atom-photon coupling is represented by $g$. Sixth term describes the interaction between the mechanical mode and the cavity photons with $\epsilon$ as the mirror-photon coupling. Last term in the Hamiltonian gives the energy due to external mechanical pump.

\begin{figure}[h]
\hspace{-0.0cm}
\includegraphics [scale=0.8]{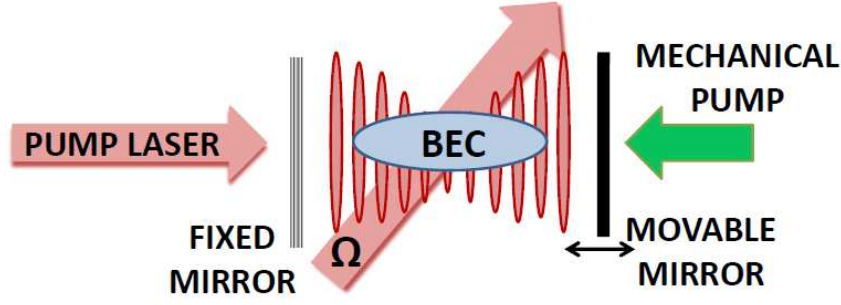}
\caption{(color online) Schematic representation of a BEC held in an optomechanical cavity, driven by a laser $\Omega$ and a pump laser in the presence of an external mechanical pump.}
\end{figure}\label{fig5}

As a convenient choice, we now rewrite the above Hamitonian in a second quantized form where the direct interaction between the atoms is characterized by the one-dimensional atomic interaction potential which is given by $u\delta(x-y)$. $u=2\hbar a_{s}\omega_{r}$ where $a_{s}$ denotes the s-wave scattering length \citep{29}. Thus the new effective Hamiltonian can be written as:

\begin{eqnarray}\label{ham3}
\hat{H} &=& \int dx \hat{\Psi}^{\dagger}(x)\left\lbrace\frac{-\hbar^{2}}{2m}\frac{\partial^{2}}{\partial x^{2}}+\frac{m\omega^{2}\hat{x}^{2}}{2}+\hbar[h(x)\hat{a}+h^{*}(x)\hat{a}^{\dagger}]+\frac{u}{2}\hat{\psi}^{\dagger}(x)\hat{\psi}(x)\right\rbrace \hat{\Psi}(x)+\hbar(\Delta-i\kappa)\hat{a}^{\dagger}\hat{a}\nonumber \\&+& \hbar\omega_{m}\hat{b}^{\dagger}\hat{b}+\hbar\epsilon\omega_{m} \hat{a}^{\dagger}\hat{a}(\hat{b}+\hat{b}^{\dagger})+\hbar \eta_{p}(\hat{b}+\hat{b}^{\dagger})-\hbar\left\lbrace\alpha_{p}\hat{a}^{\dagger}+
\alpha_{p}^{*}\hat{a} \right\rbrace,
\end{eqnarray}

where, the last term in the Hamiltonian of Eqn.(\ref{ham3}) describes the coupling of the cavity field to that of the pump laser. $\alpha_{p}$ denotes the strength of the driving field. From now on, we study the system by rescaling to dimensionless harmonic units i.e., $\hbar=\omega=m=1$.

Now, we study the equations of motion of the system by treating the operators semiclasically. Thus the condensate field operator $\hat{\Psi}(x)$, cavity field operator $\hat{a}$ and the mechanical mode operator $\hat{b}$ are replaced by the scalar quantities $\sqrt{N}\phi(x)$, $\alpha$ and $\beta$ respectively. $N$ is the number of paricles which is related to the condensate wave function by $N=\int dx \hat{\Psi}^{\dagger}(x)\hat{\Psi}(x)$. Hence, the resulting equations of motion are given as follows:

\begin{eqnarray}\label{eq6}
i\dot{\phi}(x)=\left[\frac{-1}{2} \frac{\partial^{2}}{\partial x^{2}}+\frac{x^{2}}{2}+u'\vert\phi(x)\vert^{2}+(\tilde{h}(x)\tilde{\alpha}+\tilde{h}^{\ast}(x)\tilde{\alpha}^{\ast})\right]\phi(x),
\end{eqnarray}

\begin{eqnarray}\label{eq7}
i\dot{\tilde{\alpha}}=(\Delta-i\kappa)\tilde{\alpha}+\epsilon\omega_{m}\tilde{\alpha}(\beta+\beta^{\ast})+\int dx \vert\phi(x)\vert^{2} \tilde{h}^{\ast}(x)-\tilde{\alpha_{p}},
\end{eqnarray}

\begin{eqnarray}\label{eq8}
i\dot{\beta}=\omega_{m}\beta +N\epsilon \omega_{m} \tilde{\alpha}^{\ast}\tilde{\alpha}+\eta_{p},
\end{eqnarray}

where $\tilde{h}(x)=\sqrt{N}h(x)$, $\tilde{\alpha}=\alpha/\sqrt{N}$, $u'=uN$ and $\tilde{\alpha_{p}}=\alpha_{p}/\sqrt{N}$. Now, the equations of motion (\ref{eq6})-(\ref{eq8}) are linearized around the steady state as $\phi(x)=\phi_{0}(x)+\delta\phi(x)$, $\tilde{\alpha}=\tilde{\alpha_{s}}+\delta\tilde{\alpha}$ and $\beta=\beta_{s}+\delta\beta$ where $\phi_{0}$, $\tilde{\alpha_{s}}$ and $\beta_{s}$ are the steady state values of atomic field, cavity mode and the mechanical mode respectively. $\delta\phi(x)$, $\delta\tilde{\alpha}$ and $\delta\beta$ are the respective linearized perturbations around the steady state. Moreover, eqn.(\ref{eq6}) is the time dependent Gross-Pitaevskii-like (GP) equation such that $\phi(x)=\phi_{0}(x)$ represents the ground state of unperturbed GP Hamiltonian $H_{GP}$ defined as $H_{GP}=\frac{-\partial^{2}}{\partial x^{2}}+\frac{x^{2}}{2}+u'\vert\phi(x)\vert^{2}$. When $\phi(x)$ approaches $\phi_{0}(x)$ and by considering $\tilde{\alpha_{p}}$ to be $=\int dx \vert\phi_{0}(x)\vert^{2} \tilde{h}^{\ast}(x)$, we can find the desired steady state values as $\phi(x)=\phi_{0}(x)$, $\tilde{\alpha_{s}}=0$ and $\beta_{s}=-\eta_{p}/\omega_{m}$ using Eqns.(\ref{eq6})-(\ref{eq8}) such that $\tilde{\alpha}$ will simply decay without feeding into Eqn.(\ref{eq6}). Thus the linearized equations of motion are given as follows:

\begin{eqnarray}\label{eq9}
i\delta\dot{\phi}(x)=\tilde{h}(x)\phi_{0}(x)\delta\tilde{\alpha}+\tilde{h}^{\ast}(x)\phi_{0}(x)\delta\tilde{\alpha}^{\ast}+\left(\frac{-1}{2}\frac{\partial^{2}}{\partial x^{2}}+\frac{x^{2}}{2}+2u'\phi_{0}^{2}(x)\right)\delta\phi(x)+u'\phi_{0}^{2}(x)\delta\phi^{\ast}(x),
\end{eqnarray}

\begin{eqnarray}\label{eq10}
i\delta\dot{\tilde{\alpha}}=\int dx \phi_{0}(x)(\delta\phi(x) + \delta\phi^{\ast}(x))\tilde{h}^{\ast}(x)+\delta\tilde{\alpha}(\Delta-i\kappa-2\epsilon\eta_{p}),
\end{eqnarray}

\begin{eqnarray}\label{eq11}
i\delta\dot{\beta}=\omega_{m}\delta\beta.
\end{eqnarray}

It is convenient to apply a Bogoliubov transformation \citep{54}

\begin{equation}\label{eq12}
\delta\phi(x)=\sum_{j}(\xi_{j}u_{j}(x)+\xi^{\ast}_{j}v_{j}(x)),
\end{equation}

\begin{equation}\label{eq13}
\delta\phi^{\ast}(x)=\sum_{j}(\xi_{j}v_{j}(x)+\xi^{\ast}_{j}u_{j}(x)),
\end{equation}

where the quasiparticle coefficients $\xi_{j}$ and $\xi^{\ast}_{j}$ give the normalization condition for the function $u_{j}(x)$ and $v_{j}(x)$ as

\begin{equation}
\int dx \left\lbrace u_{i}(x)u_{j}(x)-v_{i}(x)v_{j}(x) \right\rbrace=\delta_{ij}.
\end{equation}

$u_{j}(x)$ and $v_{j}(x)$ are real as $\phi_{0}$ is assumed real. They are used as a convenient time-independent basis and are orthogonal to $\phi_{0}(x)$. All the time dependence in Eqns.(\ref{eq12}) and (\ref{eq13}) thus lie within the coefficients $\xi_{j}$ and $\xi^{\ast}_{j}$, in contrast to Refs.\citep{55,56,57}. By using the transformations (\ref{eq12}) and (\ref{eq13}) and then making the integration $\int dx\left\lbrace\delta\dot{\phi}(x)u_{j}(x)+\delta\dot{\phi}^{\ast}(x)v_{j}(x) \right\rbrace $, we endup with

\begin{eqnarray}\label{eq14}
i\sum_{j}\left(\dot{\tilde{\xi}}_{j}e^{-i\omega_{j}t}-i\omega_{j}\tilde{\xi}_{j}e^{-i\omega_{j}t} \right) &=& \int dx \left\lbrace\tilde{h}(x)\phi_{0}(x) \delta\tilde{\alpha}+\tilde{h}^{\ast}(x)\phi_{0}(x)\delta\tilde{\alpha}^{\ast} \right\rbrace(u_{j}(x)+v_{j}(x))+\sum_{j} \tilde{\xi}_{j}e^{-i\omega_{j}t}A_{j}\nonumber \\&+& \sum_{j}\tilde{\xi}_{j}^{\ast}e^{i\omega_{j}t}B_{j},
\end{eqnarray}

where, $\tilde{\xi}_{j}=e^{i\omega_{j}t}\xi_{j}$ with $A_{j}$ and $B_{j}$ defined as:

\begin{eqnarray}
A_{j}=\int dx\left\lbrace\left(\frac{-1}{2}\frac{\partial^{2}}{\partial x^{2}}+\frac{x^{2}}{2}+u'\phi_{0}^{2}(x) \right)\left(u_{j}^{2}(x)+v_{j}^{2}(x) \right)+u'\phi_{0}^{2}(x)(u_{j}(x)+v_{j}(x))^{2}\right\rbrace ,
\end{eqnarray}

\begin{eqnarray}
B_{j}=\int dx\left\lbrace\left(\frac{-1}{2}\frac{\partial^{2}}{\partial x^{2}}+\frac{x^{2}}{2}+u'\phi_{0}^{2} \right)(2u_{j}(x)v_{j}(x))+u'\phi_{0}^{2}(x)(u_{j}(x)+v_{j}(x))^{2}\right\rbrace .
\end{eqnarray}

After the adiabatic elimination (assuming $\kappa>>\mid\xi_{j}\chi_{j}\mid$), Eqn.(\ref{eq7}) becomes:

\begin{eqnarray}
\delta{\tilde{\alpha}}=\frac{-\sum_{j}\chi_{j}^{\ast}(\tilde{\xi}_{j}e^{-i\omega_{j}t}+\tilde{\xi}_{j}^{\ast}e^{i\omega_{j}t})}{\Delta-2\epsilon\eta_{p}-i\kappa},
\end{eqnarray}

where the coefficient $\chi_{j}$ is defined as \citep{53}:

\begin{eqnarray}\label{eq15}
\chi_{j}=\int dx \left[u_{j}(x)+v_{j}(x) \right]\tilde{h}(x)\phi_{0}(x).
\end{eqnarray}

By substituting eqn.(\ref{eq15}) into eqn.(\ref{eq14}) and appropriately applying the rotating wave-approximation (RWA) such that $2\omega_{j}$ should therefore describe the fastest time scale, we get the damping equation:

\begin{eqnarray}
\dot{\tilde{\xi}}_{j}'=\frac{2i(\Delta-2\epsilon\eta_{p})\mid\chi_{j}\mid^{2}
\tilde{\xi}_{j}'}{\left[(\Delta-2\epsilon\eta_{p})^{2}+\kappa^{2} \right] }.
\end{eqnarray}

\begin{figure}[h]
\hspace{-0.0cm}
\begin{tabular}{cc}
\includegraphics [scale=0.80]{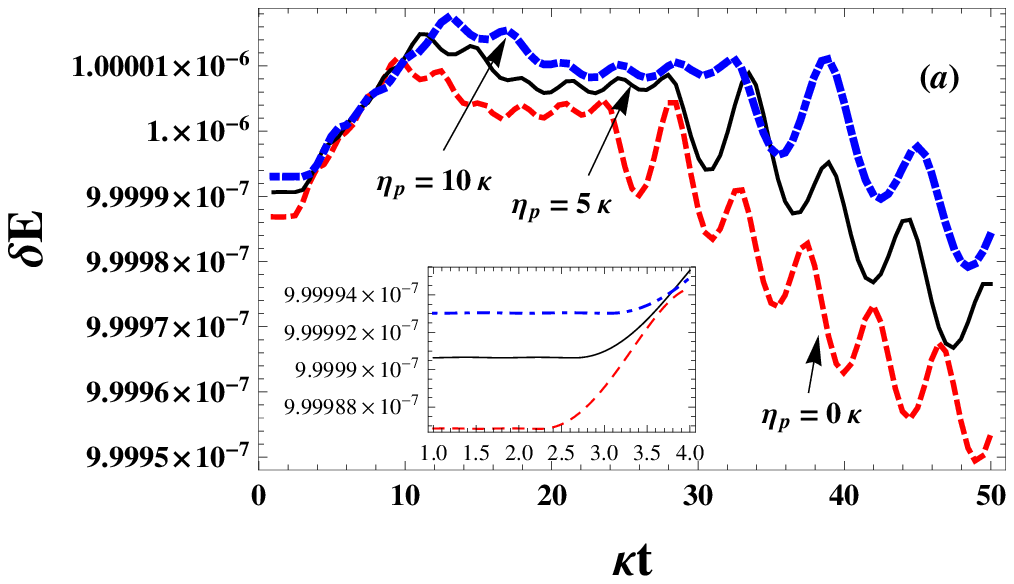}& \includegraphics [scale=
0.80]{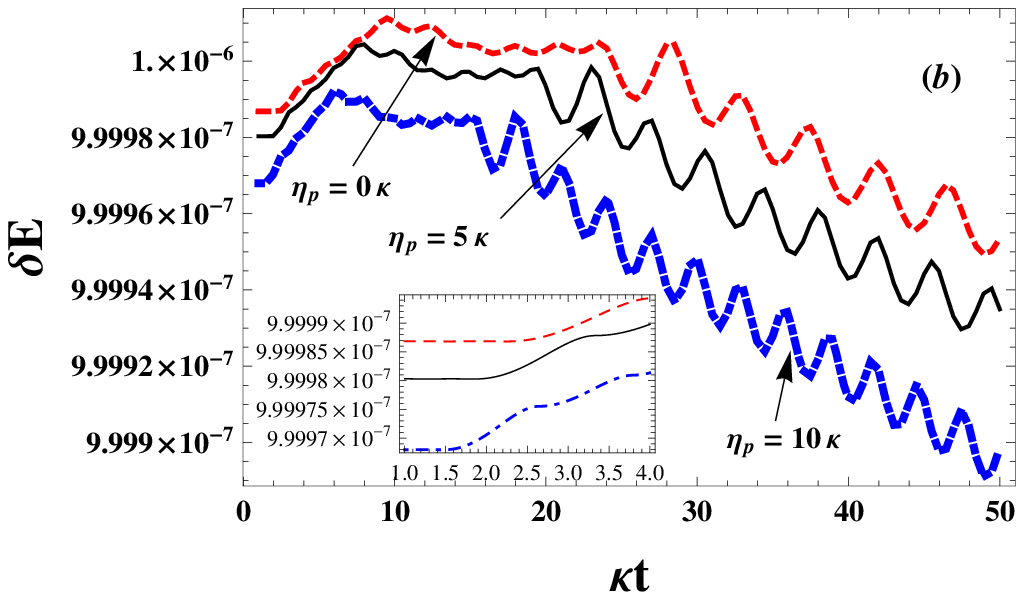}\\
\end{tabular}
\caption{(color online) Plot of energy of a Bose-Einstein Condensate as a function of time in the absence of mechanical pump (dashed line) and for two different values of mechanical pump frequency with $\eta_{p}=5$ (solid line) and $\eta_{p}=10$ (dot-dashed line). Figs.5(a) and 5(b) show the variation of energy with time for negative detuning ($\Delta=-5$) and positive detuning ($\Delta=5$) respectively. Parameters used are: $\epsilon=0.1$, $\mid\chi_{l}\mid^{2}=0.13$ and $\omega_{l}$=1. All frequencies are in the units of cavity decay rate $\kappa$.}
\label{fig6}
\end{figure}

Here $\tilde{\xi}_{j}'=e^{i(A_{j}-\omega_{j})t}\tilde{\xi}_{j}$. We now numerically examine the damping in Bose-Einstein Condensate via the change in energy $\delta E$ given by $\delta E=\sum_{j}\omega_{j}\mid\tilde{\xi}_{j}'\mid^{2}$ using Mathematica 9.0. We have $\delta E=\omega_{l}\mid\tilde{\xi_{l}'}\mid^{2}$ if only $\tilde{\xi}_{l}'\neq0$ initially. It is plotted against time in the absence of mechanical pump (dashed line) and for two different values of mechanical pump frequency with $\eta_{p}=5\kappa$ (solid line) and $\eta_{p}=10\kappa$ (dot-dashed line) (shown in fig.(6)). It shows that the energy of the condensate initially increases and then decreases with time. Fig.6(a) shows the variation in energy of BEC with time for negative detuning $\Delta=-5\kappa$. It depicts an increase in the condensate energy with time in the presence of mechanical pump. Fig.6(b) represents the energy variation of BEC with time for positive detuning $\Delta=5\kappa$. It shows the exactly opposite behaviour of BEC to that for the case of negative detuning. There is a significant decrease in the condensate energy with increase in $\eta_{p}$, thus damping the condensate. Hence, the figure clearly illustrates that higher mechanical pump frequency enhances the damping of the condensate atoms, thereby, resulting in the cooling of condensate. The quasiparticle excitations can be individually targeted in the case of a finite temperature BEC, resulting in the cooling of BEC. This implies that the mechanical pump can alter the dynamics of the condensate atoms significantly. Thus, by appropriately choosing the cavity-laser detuning and mechanical pump frequency, we have illustrated how they can in principle be used to produce extremely cold condensates. Moreover, the change in condensate energy with time by varying the mechanical pump frequency shows a coherent energy exchange between the three modes (i.e., the condensate, cavity and mirror modes). In case of positive cavity-laser detuning, the decrease in energy of the condensate atoms with time by increasing the frequency of mechanical pump cleary depicts the energy transfer to the other two modes (i.e., the cavity and mirror modes) (see fig.6(b)). This implies that, depending on whether the cavity-laser detuning is negative or positive, the coherent energy exchange between the three modes can be controlled by $\eta_{p}$.

Hence, the influence of this external force on the mirror motion results in the shifting of bifurcation point and can simultaneously change the energy of BEC atoms. However, note that the mechanical pump considered here should be chosen in such a way that the overall cooling rate must be higher than the resulting heating since some spontaneous emission is inevitable. The BEC temperature can also be affected by the atomic spontaneous emission \citep{59}.

We now discuss the experimental parameters to demonstrate that the dynamics investigated here are within the experimental reach. The mechanical mode in an optomechanical system may have frequency varying from $2\pi \times 100$Hz \citep{60}, $2\pi \times 10$kHz \citep{61}, to $2\pi \times 73.5$MHz \citep{62}. The corresponding damping rate of the movable mirror can be varied from $2\pi \times 10^{-3}$Hz \citep{60}, $2\pi \times 3.22$Hz \citep{61}, to $2\pi \times 1.3$kHz \citep{62}. The intracavity field can have decay rate $\kappa=$ $2\pi \times 1.3$MHz \citep{17} ($2\pi \times 0.66$MHz \citep{18}) interacting with the cloud of BEC that may have a coherent coupling strength of $2\pi \times 10.9$MHz \citep{17} ($2\pi \times 14.4$MHz \citep{18}). This is significantly larger than the cavity damping rate $\kappa$, thus, placing the system firmly in a regime where the Hamiltonian dynamics dominate. The mirror-photon coupling rate is $2\pi \times 2.0$MHz. The loss of photons through the cavity mirrors can be minimized by using high-finesse optical cavities in order to have strong atom-field coupling. It could also be possible to realize the critical regime of the Dicke model with just a few atoms in a regime of strong-coupling cavity quantum electrodynamics \citep{63}.

\section{Conclusion}

In conclusion, we have analyzed the semi-classical steady states for the optomechanical Dicke model in the thermodynamic limit, which demonstrate the existence of quantum phase transition (onset of self-organization) at the critical value of atom-photon coupling strength. We have also shown that the final phase of BEC held within the optomechanical cavity becomes more organized by increasing the mirror-photon coupling. We further made a semi-classical steady state analysis of the influence of an external mechanical pump (external force on the movable mirror) on the quantum phase transition. We found a shift in the critical atom-cavity field coupling strength to a lower value in the presence of external mechanical pump. This system could also be used as a new quantum device in order to measure weak forces. In addition, we investigated the effect of an external mechanical pump on the damping of a BEC confined in an optomechanical cavity via the energy. It is observed that, depending on cavity-laser detuning, the condensate energy can be increased or decreased with time by varying the mechanical pump frequency. The system also involves coherent energy exchange between the three different modes (condensate,cavity and mirror modes), which can be controlled by mechanical pump frequency. Thus, the external mechanical pump can be used as a new handle to change the critical phase transition point and to produce the extremely cold condensates. It provides a systematic control of the system by appropiately choosing the mechanical pump frequency.

\section{Acknowledgements}

Neha Aggarwal and A. Bhattacherjee acknowledge financial support from the Department of Science and Technology, New Delhi for financial assistance vide grant SR/S2/LOP-0034/2010.

\end{document}